\documentstyle{aipproc2}

\def\ga{\gamma}

\def\si{\sigma}

\def\om{\omega}

\def\De{\Delta}

\def\fr#1#2{{{#1} \over {#2}}}

\def\half{{\textstyle{1\over 2}}}
\def\frac#1#2{{\textstyle{{#1}\over {#2}}}}

\def\lsim{\mathrel{\rlap{\lower4pt\hbox{\hskip1pt$\sim$}}
    \raise1pt\hbox{$<$}}}
\def\gsim{\mathrel{\rlap{\lower4pt\hbox{\hskip1pt$\sim$}}
    \raise1pt\hbox{$>$}}}
\def\sqr#1#2{{\vcenter{\vbox{\hrule height.#2pt
         \hbox{\vrule width.#2pt height#1pt \kern#1pt
         \vrule width.#2pt}
         \hrule height.#2pt}}}}

\newcommand{\beq}{\begin{equation}}
\newcommand{\eeq}{\end{equation}}
\newcommand{\bea}{\begin{eqnarray}}
\newcommand{\eea}{\end{eqnarray}}
\newcommand{\rf}[1]{(\ref{#1})}

\begin{document}

\begin{flushright}
{IUHET 394\\}
{COLBY 98-06\\}
{August 1998\\}
\end{flushright}

\title{Testing CPT and Lorentz Symmetry with\\
Protons and Antiprotons in  Penning Traps\footnote
{\footnotesize Presented by N.R. 
at the 1998 Conference on Trapped Charged Particles
and Fundamental Physics,
Pacific Grove, California, August-September 1998}
}

\author{Robert Bluhm$^a$, V. Alan Kosteleck\'y$^b$, 
and Neil Russell$^b$}
\address{$^a$Physics Department, Colby College, Waterville, ME, 04901 U.S.A.\\
\smallskip
$^b$Physics Department, Indiana University,
Bloomington, IN, 47405 U.S.A.}

%\lefthead{LEFT head}
%\rig436thead{RIGHT head}
\maketitle

\begin{abstract}
A theoretical analysis is performed 
of Penning-trap experiments comparing protons and antiprotons
to test CPT and Lorentz symmetry
through measurements of anomalous magnetic moments
and charge-to-mass ratios.
Possible CPT and Lorentz violations 
arising at a fundamental level
are treated in the context of a general extension 
of the standard  model of particle physics
and its restriction to quantum electrodynamics.
In a suggested experiment
measuring anomaly frequencies
a bound on CPT violation 
of $10^{-23}$ 
for  a relevant figure of merit 
is attainable.
Experiments comparing 
cyclotron frequencies 
are sensitive within this theoretical framework 
to different kinds of Lorentz violation
that preserve CPT.
Constraints could be obtained on one figure of merit
at $10^{-24}$ 
and on another in 
a related experiment with 
$H^-$ ions and antiprotons
at the level of 
$10^{-25}$.
\end{abstract}

\section*{INTRODUCTION}

The standard model of particle physics 
is symmetric under the discrete transformation
CPT
\cite{cpt}.
A consequence of this is 
the equality of various 
experimentally measurable quantities.
In particular,
the charge-to-mass ratio of the proton
should equal that of the  antiproton,
and the gyromagnetic ratios
of these two particles
should be equal.
Experiments in Penning traps
can test this symmetry to high precision.
Measurements of proton and antiproton cyclotron frequencies
using Penning traps
allow a comparison of their charge-to-mass ratios
\cite{gg1},
producing the bound
\beq
r_{q/m}^p \equiv |\left[ (q_p/m_p)
- (q_{\overline{p}}/m_{\overline{p}})
\right]/(q/m)_{\rm av}|
\lsim
1.5 \times 10^{-9}
\quad .
\label{rqmp}
\eeq

In the present work,
we analyze past and future experiments 
on protons, antiprotons and hydrogen ions 
confined within a Penning trap.
The theoretical  framework is 
a CPT- and Lorentz-violating
extension of the standard model
\cite{ck}.
Since the dominant interactions are electromagnetic,
we consider the pure-fermion sector 
of the extension
of quantum electrodynamics
emerging as a limit 
of the general standard-model extension.

Our primary goal is to consider the sensitivity
of Penning-trap experiments 
to possible CPT- and Lorentz-violating effects
in the extension of quantum electrodynamics.
We investigate the relevance
of the conventional figures of merit 
as measures of CPT violation.
In some cases,
more suitable figures of merit and corresponding experiments
are suggested.
Estimates are also made of bounds 
accessible to experiments with existing technology.

\section*{THEORY}

The framework for the extension of 
the SU(3)$\times$SU(2)$\times$U(1) standard model 
and quantum electrodynamics
originates from the idea of spontaneous breaking 
of CPT and Lorentz symmetry
in a more fundamental model such as string theory
\cite{kp,ks}.
The standard-model extension lies within 
the context of conventional quantum field theory
and appears to preserve desirable features
of the standard model, including gauge invariance,
power-counting renormalizability, 
and microcausality.

At the level of the standard model,
protons and antiprotons are composite particles
formed as bound states of quarks and antiquarks,
respectively.
Possible CPT- and Lorentz-violating effects
in the standard-model extension
appear as perturbations involving the basic fields 
\cite{ck}.
A distinct set of quantities
is assigned to each quark flavor,
and suitable combinations of these determine
the CPT- and Lorentz-violating features of the proton. 
For our present investigation
of protons and antiprotons in a Penning trap,
it suffices to work within
the usual effective theory 
in which the protons and antiprotons are regarded
as basic fermions described 
by a four-component Dirac quantum field 
with dynamics governed by a minimally coupled lagrangian.
Based on the quantities for 
the fundamental particles,
we introduce effective quantities
controlling possible CPT- and Lorentz-breaking
effects for the proton.
The lagrangian is taken to be the standard one
for proton-antiproton quantum electrodynamics
but extended to include 
possible small CPT- and Lorentz-violating terms. 
Further details are given in 
Ref.\ \cite{bkr2}.

For the Penning-trap experiments of interest,
the dominant contributions to the energy spectrum
arise from the interaction of the proton or antiproton
with the constant magnetic field of the trap.
The quadrupole electric and other fields 
generate smaller effects.
In a perturbative calculation,
the dominant corrections 
due to CPT- and Lorentz-violating effects
can therefore be obtained by considering
a constant uniform magnetic field only. 
Since the signals of interest are energy-level shifts
rather than transition probabilities,
it suffices to use 
relativistic Landau-level wave functions
as the unperturbed basis set
and to calculate within first-order perturbation theory.
However,
the unperturbed energy levels  
must be taken as the Landau levels 
shifted by an anomaly term and other corrections.

To lowest order in the fine-structure constant,
the perturbative hamiltonian 
$\hat H_{\rm pert}^p$ for a proton of mass $m$ is 
\bea
\hat H_{\rm pert}^p &=& a^p_\mu \ga^0 \ga^\mu 
- b^p_\mu \ga_5 \ga^0 \ga^\mu - c^p_{0 0} m \ga^0
- i (c^p_{0 j} + c^p_{j 0})D^j 
+ i (c^p_{0 0} D_j - c^p_{j k} D^k) \ga^0 \ga^j
\nonumber \\
&&
- d^p_{j 0} m \ga_5 \ga^j + i (d^p_{0 j} + d^p_{j 0}) D^j \ga_5
+ i (d^p_{0 0} D_j - d^p_{j k} D^k) \ga^0 \ga_5 \ga^j
+ \half H^p_{\mu \nu} \ga^0 \si^{\mu \nu} \quad ,
\label{Hint}
\eea
where $D_\mu$ is the appropriate covariant derivative.
The quantities 
$a_\mu^p$, $b_\mu^p$, $H_{\mu \nu}^p$, 
$c_{\mu \nu}^p$, $d_{\mu \nu}^p$
are effective couplings for the proton and antiproton
arising in the standard-model extension.
We note that
the terms involving
$a^p_\mu$, $b^p_\mu$ 
break CPT
while those involving 
$H^p_{\mu \nu}$, $c^p_{\mu \nu}$, $d^p_{\mu \nu}$ 
preserve it,
and that Lorentz invariance is broken by all five terms.

For the antiproton,
the hamiltonian 
can be found via charge conjugation.
However, 
it should be taken into account that 
experimental procedures
for replacing particles with antiparticles in Penning traps 
typically reverse the electric field 
but leave unchanged the magnetic field.

We denote the 
proton energy levels without 
CPT- and Lorentz-violating perturbations
by $E^p_{n,s}$,
where $n$ is the principal quantum number
and $s$ is the spin.
The proton cyclotron and anomaly frequencies 
are defined as
\beq
\om_c = E_{1,+1}^{p} - E_{0,+1}^{p}
\quad ,
\quad\quad
\om_a = E_{0,-1}^{p} - E_{1,+1}^{p}
\quad .
\label{omdefsp}
\eeq
From the CPT theorem, 
these frequencies have the same values
as those of the antiproton.

We can calculate energy corrections
in first-order perturbation theory
for the CPT- and Lorentz-breaking couplings.
The corrected cyclotron and 
anomaly frequencies can then be found.
Calculations at leading order in 
the CPT- and Lorentz-breaking quantities,
in the electromagnetic fields,
and in the fine-structure constant
give the results
\beq
\om_c^{p} = \om_c^{\bar p} \approx 
(1 - c_{00}^p - c_{11}^p - c_{22}^p) \om_c
\quad ,
\label{wcp}
\eeq
\beq
\om_a^{p} \approx \om_a
+ 2 b_3^p - 2 d_{30}^p m_p - 2 H_{12}^p
\quad , \qquad 
\om_a^{\bar p} \approx \om_a
- 2 b_3^p - 2 d_{30}^p m_p - 2 H_{12}^p
\quad ,
\label{wap}
\eeq
for the modified frequencies.
In these expressions,
$\om_c$ and $\om_a$ are the unperturbed frequencies
of Eq.\ \rf{omdefsp}.

\section*{ANOMALOUS MAGNETIC MOMENTS}

Currently,
the antiproton magnetic moment 
is known to a precision of 
3 parts in $10^3$
from experiments with exotic atoms
\cite{kriessl}.
In principle,
the anomalous magnetic moments of
protons and antiprotons could be measured in Penning traps.
A comparison of the experimental ratios
$2 \om_a^p / \om_c^p$ and
$2 \om_a^{\bar p} / \om_c^{\bar p}$ 
would then provide a sharp test of CPT and Lorentz violation.
The possibility of such experiments 
has received some attention
in the literature
\cite{hw,qg}.

The sensitivity 
of possible future $g-2$ experiments
to CPT and Lorentz violation
can be investigated 
using the present theoretical framework.
We find the proton-antiproton differences
at leading order
for the cyclotron and anomaly frequencies are 
\beq
\De \om_c^p \equiv \om_c^{p} - \om_c^{\bar p} = 0
\quad , \qquad
\De \om_a^p  \equiv \om_a^{p} - \om_a^{\bar p} = 4 b_3^p
\quad .
\label{delwap}
\eeq

The leading-order signal for CPT breaking 
is thus an anomaly-frequency difference.
Denoting the exact physical energy levels 
with possible CPT violation
by ${\cal E}^p_{n,s}$ and  ${\cal E}^{\bar p}_{n,s}$, 
the corresponding figure of merit providing a well-defined
measure of the violation is 
\beq
r^p_{\om_a}
\equiv \fr{|{\cal E}_{n,s}^{p} - {\cal E}_{n,-s}^{\bar p}|} 
{{\cal E}_{n,s}^{p}}
\quad ,
\label{rp}
\eeq
where the weak-field, zero-momentum limit is understood.
We find
\beq
r^p_{\om_a}
\approx |\De \om_a^p| / 2 m_p \approx |2 b_3^p|/m_p
\quad 
\label{rp2}
\eeq
within the present theoretical framework.

Assuming an experiment could measure
$\om_a^p$ and $\om_a^{\bar p}$
with a resolution 
of the order of 
2$\pi\times$(1 Hz)
and assuming equality of $\om_c^p$, $\om_c^{\bar p}$ 
is observed to one part in $10^8$,
a bound of $|b_3^p| \lsim 10^{-15}$ eV
becomes possible.
The corresponding estimated bound 
on the figure of merit $r^p_{\om_a}$ is
\beq
r^p_{\om_a}
\lsim 10^{-23}
\quad .
\label{rp3}
\eeq

The estimate shows the 
promise held by this type of experiment
to tightly bound CPT violation in a baryon system.
The standard-model extension 
has also been applied to 
neutral mesons
\cite{k98}
and leptons
\cite{bkr}.

Measurements of diurnal variations in the anomaly frequency 
could also place bounds on 
a combination of couplings in the standard-model extension.
An estimate of one part in $10^{-21}$ 
has been made 
for a suitable figure of merit
\cite{bkr2}.

\section*{CHARGE-TO-MASS RATIOS}
 
Penning-trap experiments 
confining single protons and antiprotons
can provide precision
comparisons of their cyclotron frequencies
\cite{gg1},
yielding the limit 
$|\De\om_c^p| / \om_c^p \lsim 10^{-9}$.
Equation \rf{rqmp} gives 
the corresponding conventional figure of merit
$r_{q/m}^p$
and its bound. 

Within the present theoretical framework,
the perturbed proton and antiproton cyclotron frequencies
are given in Eq.\ \rf{wcp}.
Both
are independent of leading-order CPT-violating quantities.
As the cyclotron frequencies are unshifted
even if CPT is broken,
a comparison of these frequencies 
would represent an inappropriate measure of CPT violation
in the context of the present theory.
For example,
the figure of merit $r_{q/m}^p$ in Eq.\ \rf{rqmp},
which is proportional to
the frequency difference $\De \om_c^p$,
may vanish even though 
explicit CPT violation
occurs in the standard-model extension.

The effect of 
the Lorentz-breaking but CPT-preserving couplings
is to induce identical shifts 
in the proton and antiproton cyclotron frequencies.
This indicates that the frequency difference $\De \om_c^p$
would also be an inappropriate measure of Lorentz violation
in the present theoretical context.

Another possible experimental signal is 
the occurrence of diurnal variations
in the cyclotron frequencies,
which could be induced by the Earth's rotation 
during an experiment.
Such variations would arise in the present 
standard-model extension
from the dependence of the cyclotron frequencies on
the components $|c_{11}^p + c_{22}^p|$ of $c_{\mu \nu}^p$.

A suitable theoretical figure of merit 
can be introduced
by defining for the proton 
\beq
\De^{p}_{\om_c} \equiv 
\fr {|{\cal E}_{1,-1}^{p} - {\cal E}_{0,-1}^{p}|}
{{\cal E}_{0,-1}^{p}} 
\quad .
\label{Delpdiurnalom}
\eeq
It is the amplitude 
$r^p_{\om_c, \rm diurnal}$
of periodic fluctuations in $\De^{p}_{\om_c}$.
We find
\beq
r^p_{\om_c, \rm diurnal} 
\approx {|c_{11}^p + c_{22}^p| \om_c}/{m_p} 
\label{dclimanp}
\eeq
in the comoving Earth frame.
The appearance of $\om_c$ implies 
that the value of this figure of merit 
depends on the magnetic field.

A crude upper bound on 
$r^p_{\om_c, \rm diurnal}$
can be obtained from the data in Ref.\ \cite{gg1},
which represent alternate measurements 
of proton and antiproton cyclotron frequencies 
$\om^p_c$, $\om^{\bar p}_c$
over a 12-hour period.
The slow drifts in these frequencies 
are confined to a band of width about
$2\pi\times$(2 Hz).
This suggests a bound on a possible diurnal variation
in $r^p_{\om_c, \rm diurnal}$
arising from the contribution proportional to 
$|c_{11}^p + c_{22}^p|$,
given by
\beq
r^p_{\om_c, \rm diurnal} \lsim 10^{-24}
\quad .
\label{ccp}
\eeq
Diurnal fluctuations in
the antiproton cyclotron frequency could be 
analyzed similarly.

\section*{EXPERIMENTS WITH HYDROGEN IONS}

The precision of proton-antiproton 
cyclotron-frequency comparisons 
is limited by the need to reverse the electric field
each time the other species is loaded in the trap
\cite{gg1}.
A recent experiment by Gabrielse and coworkers
\cite{gg}
has addressed this issue 
by comparing antiproton cyclotron frequencies
with those of an $H^-$ ion instead of a proton.
The electric field is fixed throughout the experiment,
and 
the magnetic-field variation between measurements is reduced 
due to the rapid interchange possible
between simultaneously trapped hydrogen ions and antiprotons.
The expected theoretical value of the difference
$\De \om_c^{H^-}
\equiv\om_c^{H^-} - \om_c^{\bar p}$
can be obtained in the context of conventional quantum theory 
using known values of the electron mass
and the $H^-$ binding energy.
Comparison of this theoretical value
with the experimental result for 
$\De \om_c^{H^-}$
is expected to provide a symmetry test
with a ten-fold improvement on the previous 
test
\cite{gg1}.

The theoretical analysis of this experiment
within the present theoretical framework
requires a description of the electromagnetic
interactions of the hydrogen ion in a Penning trap 
in the presence of possible CPT and Lorentz violation.
The hydrogen ion is treated  as a charged composite fermion
of mass $m_{H^-}$, 
so its electromagnetic interactions can be discussed 
within an effective spinor electrodynamics
producing a modified hamiltonian 
of the form 
\rf{Hint},
but with 
a different set of 
CPT- and Lorentz-violating couplings.
The modified cyclotron frequency is then calculated as for 
the proton-antiproton case.
All the effective CPT- and Lorentz-breaking couplings
for a hydrogen ion are determined 
by appropriate combinations of the corresponding quantities 
for its constituent proton and electrons.
Lowest-order perturbation theory can be used to find
approximations to these relationships,
giving expressions involving the proton-antiproton couplings
as well as electron-positron couplings
$a^e_\mu$,
$b^e_\mu$,
$H^e_{\mu\nu}$,
$c^e_{\mu\nu}$,
and
$d^e_{\mu\nu}$.

Subject to the approximations above,
the component $\De \om_{c,\rm th}^{H^-}$ 
of $\De \om_c^{H^-}$
that is determined theoretically to arise 
purely from CPT- and Lorentz-violating effects
is
\beq
\De \om_{c,\rm th}^{H^-}
\approx
(c_{00}^{p} + c_{11}^{p} +c_{22}^{p}) 
( \om_c - \om_c^{H^-} )
- \fr{2m_e}{m_p} 
(c_{00}^{e} + c_{11}^{e} + c_{22}^{e}
-c_{00}^{p} - c_{11}^{p} -c_{22}^{p}) 
\om_{c}^{H^-} 
\quad .
\label{delwth}
\eeq
Again, $\om_c$ is the proton-antiproton
cyclotron frequency in the absence of 
CPT or Lorentz perturbations.
This result implies
that in the context of this theory
the experiment
constrains a combination of 
Lorentz-violating but CPT-preserving quantities,
including $c_{00}^e$ and $c_{00}^p$.
The latter would be inaccessible through the 
experiments with cyclotron or anomaly frequencies.
In addition,
this experiment does not look
for diurnal variations in the cyclotron frequency,
which means potential systematics
associated with diurnal field drifts are eliminated.

The definition of a model-independent figure of merit
follows from considerations similar to those 
leading to the figures of merit defined above.
We define the quantity
\beq
\De^{H^-}_{\om_c} \equiv 
\fr {|{\cal E}_{1,-1}^{H^-} - {\cal E}_{0,-1}^{H^-}|}
{2{\cal E}_{0,-1}^{H^-}} 
- \fr {|{\cal E}_{1,-1}^{\bar p} - {\cal E}_{0,-1}^{\bar p}|}
{2{\cal E}_{0,-1}^{\bar p}} 
\quad .
\label{DelHomc}
\eeq
As given here,
$\De^{H^-}_{\om_c}$ is nonzero 
even if CPT and Lorentz symmetry is preserved.
To arrive at a measure that vanishes in the exact symmetry limit,
we remove from the hydrogen-ion terms in $\De^{H^-}_{\om_c}$
the conventional contributions arising from the differences
between the $H^-$ ion and a proton:
the masses of the two electrons and the binding energy.
The result is an appropriate figure of merit for 
Lorentz violation,
denoted by $r^{H^-}_{\om_c}$.

Estimating a precision of 
one part in $10^{10}$ 
in measurements of the ratio 
$|\De \om_c^{H^-}|/\om_c^{H^-}$,
we estimate an experimentally attainable bound of 
$r^{H^-}_{\om_c} \lsim 10^{-25}$.
Indeed, the Gabrielse experiment 
\cite{gg}
placed a bound of 
\beq
r^{H^-}_{\om_c} \lsim 4 \times 10^{-26}
\quad
\eeq
on this figure of merit.

\section*{CONCLUSIONS}

We have used a general theoretical framework
based on an extension of the standard model
and quantum electrodynamics
to establish and investigate possible signals
of CPT and Lorentz breaking
in Penning-trap experiments 
with protons and antiprotons.
We have looked for leading-order limits  
arising from precision measurements 
of anomaly and cyclotron frequencies.

Sharp tests of CPT symmetry would be possible 
in experiments comparing anomaly frequencies.
We have introduced appropriate figures of merit
with attainable bounds of approximately 
$10^{-23}$ for a plausible experiment 
with protons and antiprotons.

In contrast,
comparative measurements of cyclotron frequencies
for protons and antiprotons are insensitive 
to leading-order effects from 
CPT breaking within the present framework.
However,
diurnal variations of cyclotron frequencies
and comparisons of cyclotron frequencies
for hydrogen ions and antiprotons are affected by different 
CPT-preserving Lorentz-violating quantities.
These experiments could generate bounds 
on various dimensionless figures of merit 
at the level of
$10^{-24}$ in the proton-antiproton system,
and $10^{-25}$ using the $H^-$-antiproton system.

\section*{ACKNOWLEDGMENTS}
 
This work is supported in part 
by the Department of Energy
under grant number DE-FG02-91ER40661 
and by the National Science Foundation
under grant number PHY-9801869.

\end{document}